\documentclass[%
 preprint, 
superscriptaddress,
 amsmath,amssymb,
 aps, physrev,
]{revtex4-2}

\usepackage{graphicx}
\usepackage{dcolumn}
\usepackage{bm}

\usepackage[colorlinks=true,urlcolor=blue,linkcolor=blue,citecolor=blue]{hyperref}

\begin{document}

\preprint{APS/123-QED}

\title{\textbf{Sliding of liquid droplets on thin viscoelastic soft layers} 
}%

\author{Menghua Zhao}
\affiliation{School of Information Mechanics and Sensing Engineering, Xidian University, Xi'an 710126, China}%
\affiliation{Université Paris Cité, CNRS, Matière et systèmes complexes, F-75013 Paris, France}
\affiliation{Sciences et Ingénierie de la Matière Molle, CNRS UMR 7615, École supérieure de physique et de chimie industrielles de la Ville de Paris, Sorbonne Université, Paris Sciences et Lettres Université, Paris 75005, France}

\author{Julien Dervaux}
\affiliation{Université Paris Cité, CNRS, Matière et systèmes complexes, F-75013 Paris, France}

\author{Tetsuharu Narita}
\affiliation{Sciences et Ingénierie de la Matière Molle, CNRS UMR 7615, École supérieure de physique et de chimie industrielles de la Ville de Paris, Sorbonne Université, Paris Sciences et Lettres Université, Paris 75005, France}

\author{François Lequeux}
\affiliation{Sciences et Ingénierie de la Matière Molle, CNRS UMR 7615, École supérieure de physique et de chimie industrielles de la Ville de Paris, Sorbonne Université, Paris Sciences et Lettres Université, Paris 75005, France}

\author{Laurent Limat}
\affiliation{Université Paris Cité, CNRS, Matière et systèmes complexes, F-75013 Paris, France}

\author{Matthieu Roché}
\email{Contact author: matthieu.roche@univ-parisdiderot.fr}
\affiliation{Université Paris Cité, CNRS, Matière et systèmes complexes, F-75013 Paris, France}


\date{\today}

\begin{abstract}
Soft substrates are deformed by liquid-vapor surface tension upon contact with liquid droplets, forming the well-known wetting ridge. This ridge dynamically propagates with the moving contact line and critically influences liquid spreading. Here, we experimentally investigate gravity-driven sliding dynamics of water droplets on vertically tilted silicone layers whose viscoleasticity is characterized by the Chasset-Thirion model with the exponent $m$. At low Bond numbers, the sliding velocity scales with droplet size as $V_S \sim D^{\frac{2}{m}}$. While in the thin-film limit, velocity exhibits a pronounced power-law dependence on nominal substrate thickness, $V_S \sim \Pi(h)^{-\frac{1}{m}}$. We rationalize these observations by quantifying viscoelastic dissipation within the soft layer and balancing it against the gravitational driving force using an energy-conservation framework. Our findings offer novel avenues for designing advanced soft coatings, anti-fouling and self-cleaning surfaces, and biomedical devices.

\end{abstract}

\maketitle


\section{\label{sec:level1}Introduction}

Soft gels find wide applications in drug delivery \cite{wang2021enhanced}, medical adhesives \cite{Zhang2025Stimuli}, soft robotics \cite{zhao2022stimuli} due to their high programmability. When in contact with a droplet, they can be deformed by the liquid-vapor surface tension in the vicinity of the contact line, forming a wetting ridge \cite{shanahan1994anomalous,carre1996viscoelastic,andreotti2020statics} of a characterized height $\gamma_{LV}/E$, where $\gamma_{LV}$ and $E$ denote the liquid-vapor surface tension and Young's modulus, respectively. While such deformation is negligible for rigid materials (less than one atom for materials of $E\sim$ GPa), it becomes significant for soft substrates ($\mathcal{O}(10) \mu$m for $E\sim \mathcal{O}(10)$ kPa). During contact line motion, the propagation of this deformation leads to energy dissipation in the substrate that can be greater than viscous dissipation in the liquid phase and profoundly influences spreading dynamics through the viscoelastic braking effect by slowing down the propagation velocity \cite{shanahan1994anomalous,carre1996viscoelastic,long1996static,karpitschka2015droplets,zhao2018geometrical}, challenging classical hydrodynamic models that only describe the flow in the liquid phase \cite{blake2006physics,snoeijer2013moving,andreotti2020statics,roche2024complexity}. The interplay between viscoelastic dissipation in the gel phase and viscous dissipation in the liquid phase introduces a complex spectrum of behaviors, while simultaneously enabling broader applications, such as directional transport \cite{zhao2018geometrical,bardall2020gradient,galatola2022lateral}, biomedical engineering\cite{zhang2021wetting,yi2022overview}, and adaptive surface design \cite{wong2020adaptive}.

Extensive studies focused on the study of the static surface deformation close to the contact line using interferometry \cite{shanahan1994anomalous,carre1996viscoelastic}, confocal microscopy\cite{pericet2008effect,style2013universal}, X ray microscopy\cite{park2014visualization} and microscopic Schlieren imaging \cite{zhao2018geometrical}, revealing the logarithmic geometry of the wetting ridge. Building upon these advances, more attention was subsequently devoted to dynamic scenarios where energy dissipation in the soft film phase was  accounted for \cite{carre1996viscoelastic,karpitschka2015droplets,zhao2018geometrical} and pertinent factors including the geometrical confinement \cite{zhao2018geometrical}, physico-chemistry \cite{hourlier2017role,hourlier2018extraction}, non-linear effects \cite{dervaux2020nonlinear,masurel2019elastocapillary}, Shuttleworth effect \cite{bain2021surface} were thoroughly addressed. However, the majority of preceding research has focused on a more idealized situation than the actual sliding process, where a droplet can deform and the substrates can undergo residual stress \cite{podgorski2001corners,oleron2024morphology,chao2024non,xue2025droplets}. 
For example, in the case of sliding on thick soft substrates driven by gravity, droplet morphology has been shown to exhibit a pronounced dependence on the Bond number and relaxation ratio \cite{oleron2024morphology}: droplets are more elongated at a higher Bond number and tend to be obtuser for a larger relaxation ratio. Such a morphology dependence is similarly recovered in alternative research context \cite{xue2025droplets}, where a droplet is forced to slide horizontally on the soft silicone substrate and the relaxation of elastocapillary ridges is additionally reported.
In addition, recent studies have demonstrated that droplets exhibit anisotropic descent from a soft, stretched substrate, displaying enhanced velocity in the stretched direction in comparison to the perpendicular direction \cite{smith2021droplets}. While in a compression case, droplets exhibit a highly non-monotonic dependence on the strain applied \cite{chao2024non}. Most those research on soft sliding of droplets has been conducted on soft films of large thickness, despite the established significance of small thickness in determining contact line dynamics \cite{zhao2018geometrical}. Furthermore, the driving force is typically difficult to control in most contexts, which complicates a precise estimation of the dissipation power. 

The present research focuses on the gravity-driven sliding of water droplets across vertically tilted soft substrates where the driving power can be precisely controlled. Sliding dynamics will be addressed by tailoring the droplet size and the thickness of the film over three orders of magnitude.
The findings of this study demonstrate that increasing the thickness of the soft layer results in enhanced energy dissipation, consequently leading to a reduction in sliding velocity. A model has been formulated by balancing the viscoelastic dissipation and the gravitational driving power to account for the experimental observations. The study reveals a scaling law that relates the droplet size and the thickness of the viscoelastic layer to the sliding velocity in the limit of small Bond number.

\section{Experimental setup}

\begin{figure}[bt!]
\centering
	\includegraphics[width=0.6\columnwidth]{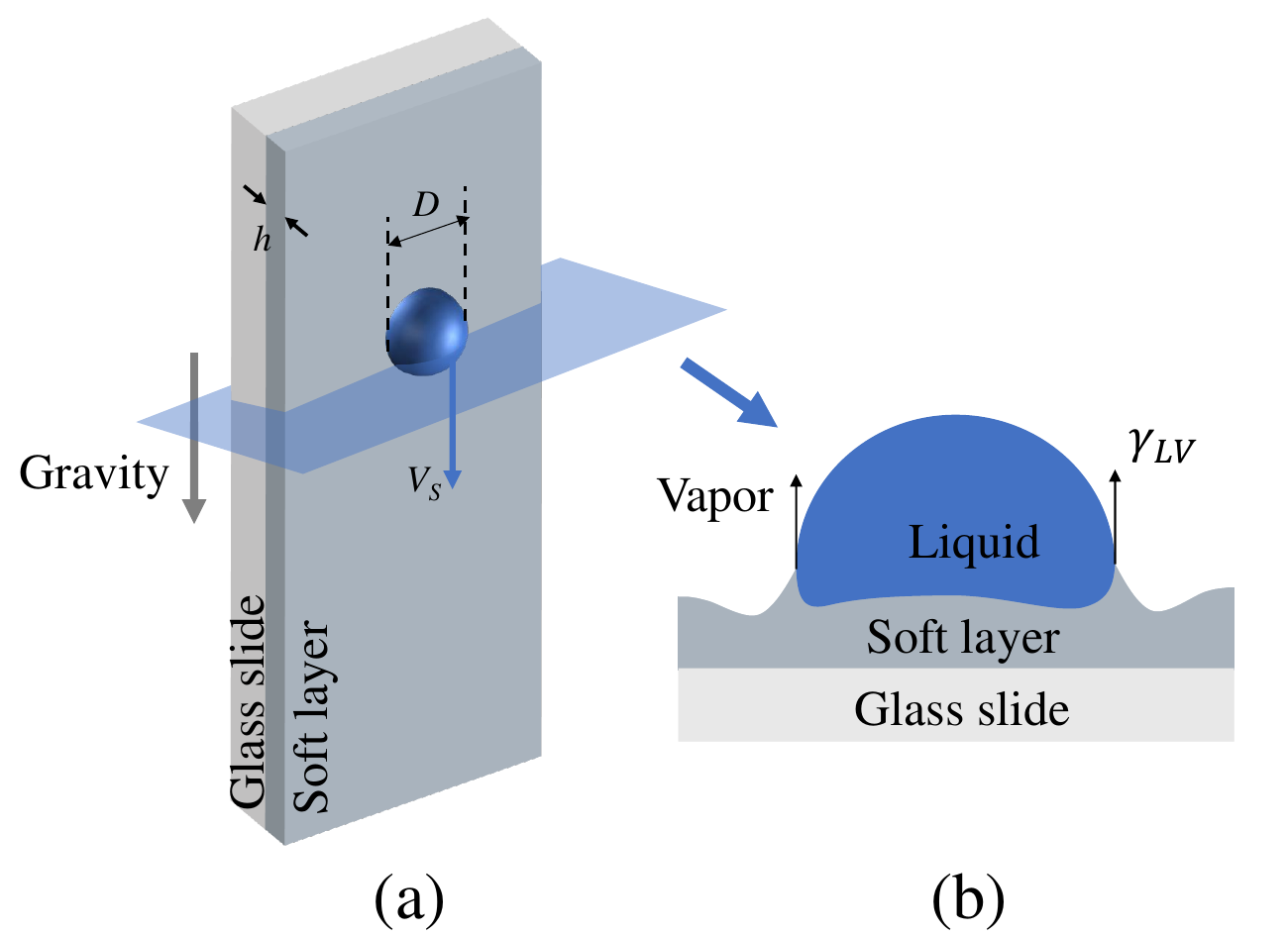}
\caption{\label{fig:Exp} Schematic of the sliding experiments. (a) 3D diagram of the water droplet sliding on the silicone gel layer along the gravity direction. The layer thickness is controlled by spin coating, ranging from $\sim \mu$m to $\sim$ mm. (b) 2D sketch of the surface deformation induced by the liquid-vapor surface tension close to the contact line.}
\end{figure}

Droplet sliding experiments were performed in a humidity-controlled chamber with soft films being bonded onto vertically tilted glass slides, establishing a 90$^{\circ}$ sliding angle (Fig.~\ref{fig:Exp}). The tilting angle was monitored by an inclinometer with a precision of $\pm$0.1$^{\circ}$. The assembly was housed in a vertically oriented, enclosed Petri dish containing additional water drops to maintain the saturation vapor pressure, thereby suppressing evaporation during sliding. To validate evaporation suppression, the smallest droplets ($\sim$1 $\mu$L) were monitored and shown to exhibit a diameter reduction from 1.25 mm to 1.23 mm over $\sim$1 hour's sliding, corresponding to a negligible relative change of 1.6$\%$. These results confirm that our experiments were conducted in a vapor-saturated environment, allowing evaporation effects to be safely neglected. To visualize and track the sliding dynamics, a LED array was employed to provide homogeneous illumination from the back side and a front view camera (DFK 23UX174, IMAGING SOURCE, Germany) was installed facing the soft film with a spatial resolution of 19 $\mu$m/pixel at the frame rate of 20 Hz. Droplet tracking was implemented by means of the correlation method with the sub-pixel technique enabling a localization uncertainty of $\pm 2\,\mu m$ \cite{zhao2012combined}. Droplet deposition was achieved by gently injecting liquid onto soft films with a micro-pipette that defines the droplet volume with a precision of $\pm 0.2\, \mu$L. The liquid employed in this study was distilled water (Milli-Q Integral; Millipore, USA), with the surface tension $\gamma_{LV}$ of 72.8\,mN$/$m and the viscosity $\eta$ of 1.0\,mPa$\cdot$s. The rheological response of the silicone layer, prepared following a protocol we published previously \cite{zhao2018geometrical}, is described by the Chasset-Thirion model, $G(\omega)=G_s(1+(i\omega\tau)^m)$. We find a shear modulus $G_s$\,=\,1.2\,kPa, a relaxation time $\tau$\,=\,8.6\,ms, and an exponent $m$\,=\,0.62.
The soft gel film was prepared by the spin coating technique and its thickness was tuned by spinning speed, allowing a continuous control of the thickness from $\sim \mu m$ to $\sim$ mm. That was quantitatively measured by a 3D Profiler (FOCAL 3D Pilot, FOCAL Nanotech) in the white light
scanning mode.

\section{Experimental observations}

\begin{figure*}[bt!]
\includegraphics[width=1\columnwidth]{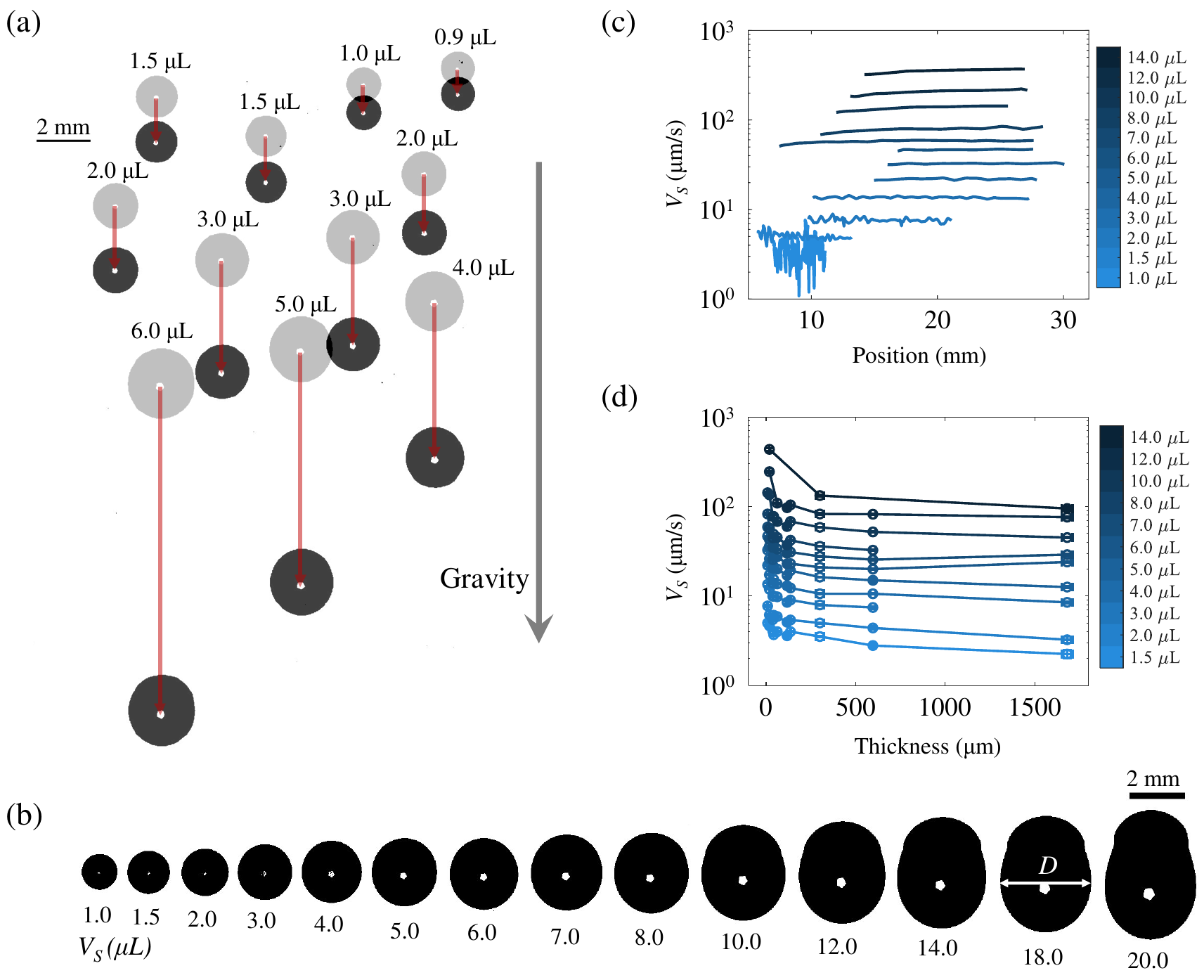}
\caption{\label{fig:Sliding}Droplets sliding on silicone gel layers.
(a) Superimposed images of droplets recorded with a time lag of 340 s. The initial droplet position (gray) and its position after 340 s (black) illustrate the sliding motion. The red arrow indicates the sliding trajectory. Film thickness is 18.5 $\mu$m.
(b) Morphology of steady sliding droplets as a function of droplet size.
(c) Sliding velocity as a function of sliding displacement. Position 0 corresponds to the initial deposition position of the first droplet.
(d) Sliding velocity as a function of the soft layer thickness.}
\end{figure*}

We first report the collective sliding behaviors on a soft film of fixed thickness in Fig.~\ref{fig:Sliding} by varying droplet size, $D$. 
Following the initial deposition of a droplet, a transient acceleration regime ensues, subsequently giving way to a steady state characterized by a constant sliding velocity, $V_S$. This transient regime happens during the very first tens of seconds and is not the primary subject of interest in this study. Fig.~\ref{fig:Sliding}(a) presents the sliding trajectories of droplets with varying sizes (0.9 $\mu$L to 6.0 $\mu$L) from 60 s to 400 s after their initial deposition. We observe that all those droplets slide down along the gravitational direction and maintain a spherical shape. Red arrows denote sliding displacements over a fixed 340\,s observation period, indicating that $V_S$ increases with droplet size. To analyze droplet morphology, two dimensionless parameters are critical: the Capillary number (Ca=$\eta V_S/\gamma_{LV}$) characterizing the predominance of the viscous force and the surface tension at the interface, and the Bond number (Bo=$(D/l_c)^2$), with $l_c$=$\sqrt{\gamma_{LV}/\rho g}$ the capillary length. The calculated Ca value ranges from $\sim$1.4$\times10^{-8}$ to $\sim$1.4$\times10^{-5}$, suggesting negligible viscous effects during sliding. 
Meanwhile, Bo values near unity highlight the importance of droplet size in shaping morphology. As shown in Fig.~\ref{fig:Sliding}(b), droplet shape evolves with volume: droplets remain spherical for $D<l_c$, while elongating in the gravitational direction at $D>l_c$. However, a precise prediction of the exact shape as a function of large Bo still remains challenging and will not be the main focus of this study. 
Inspired by the Oléron's work on the droplet mophology during the sliding \cite{oleron2024morphology}, we further estimate the relaxation ratio $R$ ($R=\gamma_{LV} \tau/\eta l_s$ with $l_s=\gamma_s/(2G_s)$) that characterizes dominance of the viscocapillary relaxation in the liquid phase and that in the solid phase \cite{dervaux2020nonlinear}. In the present study, this $R$ is a constant of 20640, indicating that the relaxation in the liquid is much faster than that in the gel phase. Interestingly, the rear of the running droplets exhibits different morphology than that reported at a smaller $R$ \cite{oleron2024morphology}: the rear thins down collectively rather than forms pinched-off pearlings at a high Bo number \cite{podgorski2001corners,oleron2024morphology}, possibly pointing to a different instability mechanism which deserves further study.

Another finding is that the sliding velocity increases with the droplet volume, consistent with prior studies \cite{podgorski2001corners,oleron2024morphology}. A simplified scaling argument supports this trend. The dissipation power during the sliding should scale with the total length of the contact line, $P\sim D$, based on the coarse assumption that the dissipation in the gel phase is uniform along the contact line. Concurrently, the driving power scales with the droplet volume, $P_d \sim D^3V_S$, provided that the pinning force during the sliding is neglected. Balancing the two yields a proportionality between the sliding velocity with the droplet size. This analysis, however, offers only a qualitative framework; quantitative predictions would require a more rigorous treatment of dissipation mechanisms. 

Focusing on the sliding velocity along the sliding trajectory shown in Fig.~\ref{fig:Sliding}(c), we conclude no significant physicochemical variations during steady-state sliding. This result contrasts with previous studies suggesting that uncrossed free chains within the silicone elastomer may migrate and coat the sliding droplets once they are in contact \cite{hourlier2017role,hourlier2018extraction}. A plausible resolution to this discrepancy lies in the high concentration of free chains within the silicone gel used here \cite{oleron2024morphology} , which may have already modified the droplet surface during the initial acceleration phase. Consequently, the steady sliding regime likely corresponds to a droplet in a "contaminated" state. Critically, this surface modification does not compromise our analysis of droplet size and thickness effects, as the droplet surface state stabilizes after the transient regime. Furthermore, as discussed later, the impact of this contamination is minor and can be attributed to a slight adjustment in surface tension rather than a fundamental alteration of interfacial dynamics.

Furthermore, the present study demonstrates a pronounced effect of the gel layer thickness on the sliding velocity, as illustrated in Fig.~\ref{fig:Sliding}(d),  which is in accordance with our previous report \cite{zhao2018geometrical}: the thinner the thickness, the faster the sliding velocity. While conventional analysis estimates driving energy from deviations in the liquid-vapor interface relative to static conditions, our system enables direct manipulation of the driving power thanks to the absence of contact line pinning on silicone gels \cite{perrin2016defects,zhao2017wetting,oleron2024morphology}. This departure from conventional approaches highlights the unique control of interfacial dynamics in our experiments. However, due to the inconformity of the contact angle and the deviation of the sliding direction from the deformation symmetric axis along the contact line, the dissipation along the contact line is expected to be position-dependent. This raises questions about the universality of existing dissipation models where the contact line moving direction is the same as the deformation symmetric axis. Systematically adapting existing dissipation models to this asymmetric regime presents an intriguing avenue for future investigation.

\section{Model and discussion}

To rationalize the aforementioned results, we move towards an analytical solution by balancing the gravity energy with the dissipated energy. Within this framework, the dissipation power of the contact line per unit length is expressed as at small spreading velocity for the silicone gel characterized by the Chasset-Thirion model \cite{zhao2017wetting,zhao2018geometrical,oleron2024morphology},

\begin{eqnarray}
\begin{split}
P_{\text{film}} \sim (\frac{\gamma_{LV}}{\gamma_S})^2G_S V_S^{m+1} \tau^{m} \Pi(h)
\label{eq:dissipation}
\end{split}
\end{eqnarray}

with
\begin{eqnarray}
\begin{split}
&\Pi(h)=\int_{0}^{\infty} \frac{|k|^{m+2}}{(k^2+\frac{G_S}{\gamma_S K(k)})^2} \,dk  
\\ &K(k)=\frac{1}{2k}[\frac{sinh(2hk)-2hk}{2h^2k^2+cosh(2hk)+1}]&   
\label{eq:dissipationNum}
\end{split}
\end{eqnarray}

Now we estimate the overall dissipation by considering multiple dissipation channels and contact line morphology. First, as shown in prior studies \cite{carre1996viscoelastic,karpitschka2015droplets,zhao2017wetting,zhao2018geometrical,oleron2024morphology,xue2025droplets}, the dissipation at the wetting ridge in the gel phase dominates over that in the liquid phase for the silicone gel used in this work where the relaxation ratio $R\gg1$. Consequently, we will exclusively analyze dissipation in the soft gel phase. In the case of small droplets at Bo$<$1 where surface tension dominates, the contact line adopts a round shape, in which case the total length of the contact line scales with $D$. For Bo$>$1, the sliding droplet elongates along the sliding direction, making the aforementioned claim questionable \cite{oleron2024morphology,xue2025droplets}. A precise description of the droplet shape on the soft substrate at Bo$>$1 remains elusive, preventing the formulation of a precise relation between the droplet size and the contact line length. In light of these difficulties, we simply assume that the total contact line length scales with $D$ across all Bond numbers. In the following, we examine the second assumption. It is important to note that Eq.~(\ref{eq:dissipation}) is only valid for a dynamic contact angle of 90$^{\circ}$, which is not applicable to a sliding droplet. The lack of a description of the three-dimensional shape of the droplet shape further hinders precise characterization of surface deformation and local dissipation along the contact line during the sliding. We thus approximate the dissipation power per unit contact line length as uniform, yielding a total dissipation power scaling as $P_{\text{film}}D$. This assumption is reasonable on the whole by potential mutual compensatory effects between deformation-induced dissipation at the advancing and receding contact line. The driving force for sliding arises solely from gravity, as pinning forces are negligible on silicone gels \cite{zhao2017wetting,oleron2024morphology}. Balancing the dissipation energy and gravitational work over a sliding duration $\Delta t$, we derive,

\begin{equation}
D \Delta t P_{\text{film}} = m g V_s \Delta t. 
\label{eq:energybalance}
\end{equation}

By injecting Eq.~(\ref{eq:dissipation}) and Eq.~(\ref{eq:dissipationNum}) into Eq.~(\ref{eq:energybalance}), we obtain the following relationship,

\begin{equation}
V_S^m \sim (\frac{\gamma_S}{\gamma_{LV}})^2 (\frac{\rho D^2 g}{G_0 \Pi(h) \tau^m}). 
\label{eq:Nondimentionless}
\end{equation}

To reach a dimensionless relation, we rearrange terms on the left side and right side and finally obtain the following scaling,

\begin{equation}
\frac{V_S\tau}{l_c} \sim (\frac{\gamma_S}{\gamma_{LV}})^{\frac{2}{m}} (\frac{\rho g}{G_0 l_c^{m-2}\Pi(h)})^{\frac{1}{m}} (\frac{D}{l_c})^{\frac{2}{m}}. 
\label{eq:Dimentionless}
\end{equation}

\begin{figure*}
\includegraphics[width=1\columnwidth]{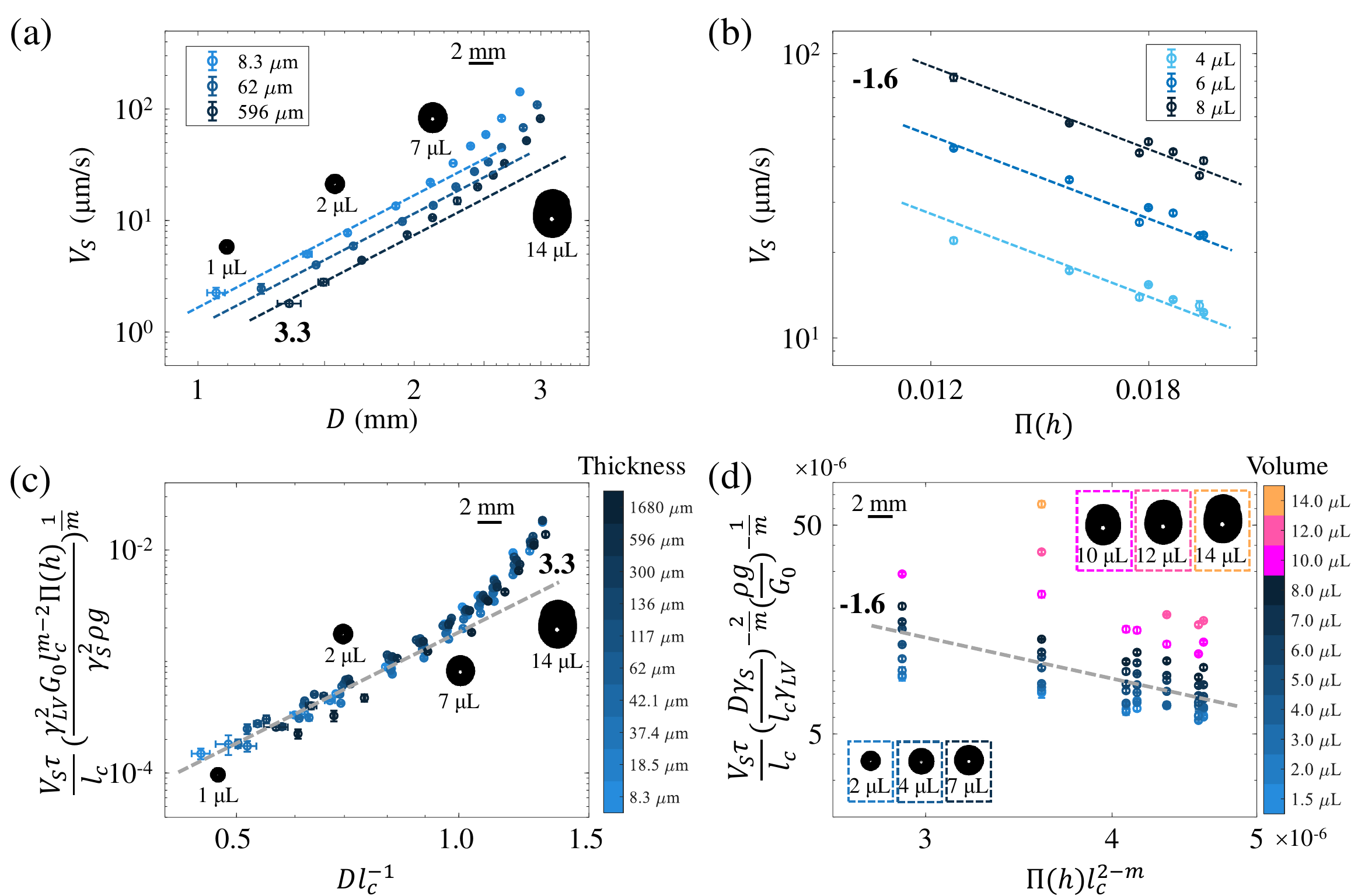}
\caption{\label{fig:Scaling}Scaling law of droplet sliding. 
(a) Sliding velocity as a function of the droplet size on soft films with different thickness: 8.3\,$\mu$m, 62\,$\mu$m, 596\,$\mu$m. The dotted line corresponds to the power law fitting with the exponent of 3.3. 
(b) Sliding velocity as a function of nominal thickness $\Pi(h)$ for droplets of different volume: 4\,$\mu$L, 6\,$\mu$L, 8\,$\mu$L. The dotted line corresponds to the power law fitting with the exponent of -1.6. 
(c) Power-law relationship between the nondimensionalized sliding velocity and the droplet size normalized by the capillary length,  $D l_c^{-1}$. The dotted gray line represents the theoretical scaling.
(d) Power-law relationship between the nondimensionalized sliding velocity and the nominal thickness of the thin film $\Pi(h)l_c^{2-m}$. The dotted gray line represents the model scaling.}
\end{figure*}

All parameters are determined experimentally, with the exception of surface tension $\gamma_{LV}$, due to uncertainties in the surface state caused by the substantial presence of free polymer chains in the silicone gel that can migrate and coat the droplet interface \cite{hourlier2017role,hourlier2018extraction,oleron2024morphology}. The thickness factor $\Pi(h)$, quantifying the thickness effect, is computed numerically. 
Fig.~\ref{fig:Scaling}(a) shows the experimental scaling relation between the droplet size and sliding velocity on thin films of different thickness. Their extensive dimensionless comparison is further demonstrated in Fig.~\ref{fig:Scaling}(c) where the dimensionless sliding velocity is associated with the normalized droplet size via the same simple scaling law, $V_S \sim D^{3.3}$. That aligns closely with our analytical solution presented by Eq.~(\ref{eq:Dimentionless}) for $D<l_c$ where $V_S \sim D^{2/m}$. Using $m$=0.62 derived from rheological measurement, the theoretical exponent $2/m$=3.2, matches the experimental value of 3.3. This agreement supports the assumptions we made in writing our model. However, for large droplets of $D>l_c$, systematic deviations from the theoretical curve emerge, likely linked to droplet elongation and morphological changes in the contact line in Fig.~\ref{fig:Sliding}(b). Notably, while advancing and receding contact angles deviate from the 90$^\circ$ reference value, their combined effects seem to compensate spatially, preserving the proportionality between total dissipation and contact line length. We further comment on the surface tension that is left as a fitting parameter. The fitting of the experimental data to the theoretical prediction yields $\gamma_{LV}$=55 mN/s, a value that closely approximates the measured value from the prior contamination test \cite{hourlier2017role}, reinforcing the consistency of the proposed model. This concordance underscores the robustness of the simplified framework in explaining the observed scaling dynamics.

For the thickness effect, we first report the experimental scaling relation between the sliding velocity and the nominal thickness, $\Pi(h)$ for three typical droplet sizes in Fig.~\ref{fig:Scaling}(b): $V_S$ scales with $\Pi(h)$ via a power law with the exponent of -1.6.
To generalize a consistent way to report our observations, we reformulate Eq.~(\ref{eq:Dimentionless}) and further show the dimensionless velocity as a function of the dimensionless thickness $\Pi(h)l_c^{2-m}$ in Fig.~\ref{fig:Scaling}(d). We reveal a consistent agreement between experiments and the model outlined in Eq.~(\ref{eq:Dimentionless}) for small droplet volume (<10\,$\mu$L). Direct comparison with $h$ is avoided, as the model focuses on $\Pi(h)$ rather than $h$ itself, though prior work \cite{zhao2018geometrical} links $\Pi(h)$ to $h$ at small $h$. Our analysis hereby extends to broader thickness ranges, revealing an experimental scaling exponent of -1.6, matching the theoretical prediction, $-1/m$=-1.6. This confirms thickness-dependent control of dissipation. Two systematic deviations merit discussion. First, nondimensionalized sliding velocity still exhibits the droplet size dependence, and such dependence seems to persist in Fig.~\ref{fig:Scaling}(d) in line with the thickness effect, in sharp contrast to the report in Fig.~\ref{fig:Scaling}(c) where the effects from $\Pi(h)$ and $D$ can be separately demonstrated. The entanglement of the two effects in Fig.~\ref{fig:Scaling}(d) suggests other second-order mechanisms in operation other than the simple model proposed here and the droplet size effect must overshadow the $\Pi(h)$ effect and other possible effects. Given the circular contact morphology, this other possible effect likely comes from the dynamic contact angle, a considerable deviation from the model assumption. Second, significant departures from theoretical predictions occur for large droplets (>8\,$\mu$L), exceeding explanations above. Given the fact of the contact line morphology dependence on the droplet volume shown in Fig.~\ref{fig:Sliding}(b), contact line morphology likely dominates here. What is interesting though is that both the dynamic contact angle and the contact line morphology trigger less the total energy dissipation compared to the case of the 90$^\circ$ sliding droplet. Current knowledge prevents full incorporation of dynamic contact angles and droplet deformation effects, necessitating future refinements.

\section{Conclusion}

In this study, we experimentally investigate the vertical sliding dynamics of water droplets on silicone gel substrates by systematically varying droplet size and gel layer thickness. Quantitative characterization of droplet morphology and sliding velocity reveals distinct controls by both parameters. For Bond numbers Bo$<$1, the contact line of sliding droplets exhibits a circle shape, while it is elongated with the increase of the Bond number at Bo$>$1. The sliding velocity increases with the droplet size by a power law with the exponent 3.3, and scales with the nominal thickness $\Pi(h)$ by a power law with exponent -1.6 at the limit of Bo$<$1. Using the Chasset-Thirion model within a linear viscoelastic framework, we formulate a dissipation model for the gel phase and balance it against gravitational driving energy. Key assumptions includes: (1) a circular contact line, (2) a fixed dynamic contact angle of 90$^\circ$, (3) constant liquid-vapor surface tension. Despite these simplifications, the model successfully reproduces the observed scaling laws between the sliding velocity and the droplet size, thickness. This research sketches a framework for the sliding dynamics by considering the energy dissipation from the gel phase, with implications for applications such as dew harvesting, self-cleaning surfaces. The following challenges include the incorporation of three-dimensional droplet morphology during sliding and dissipation mechanisms under multidirectional substrate deformations.

\newpage

\begin{acknowledgments}
This work was supported by Agence Nationale de la Recherche and Commissariat à l’Investissement d’Avenir Multiscale Modeling of Material Interfaces (MMEMI) Grant (LabEx Science and Engineering for Advanced Materials and Devices) ANR-11-LABX-086, ANR-11-IDEX05-02 and Poroelasticity and Non-linearity in Gel Wetting (GELWET) Grant ANR-17-CE30-0016.

\end{acknowledgments}

\section*{Data Availability Statement}
All data are included in the manuscript.

\appendix

\section{Rheology characterization}

\setcounter{figure}{0} 
\renewcommand{\thefigure}{A\arabic{figure}} 

\begin{figure}[h]
\centering
	\includegraphics[width=0.6\columnwidth]{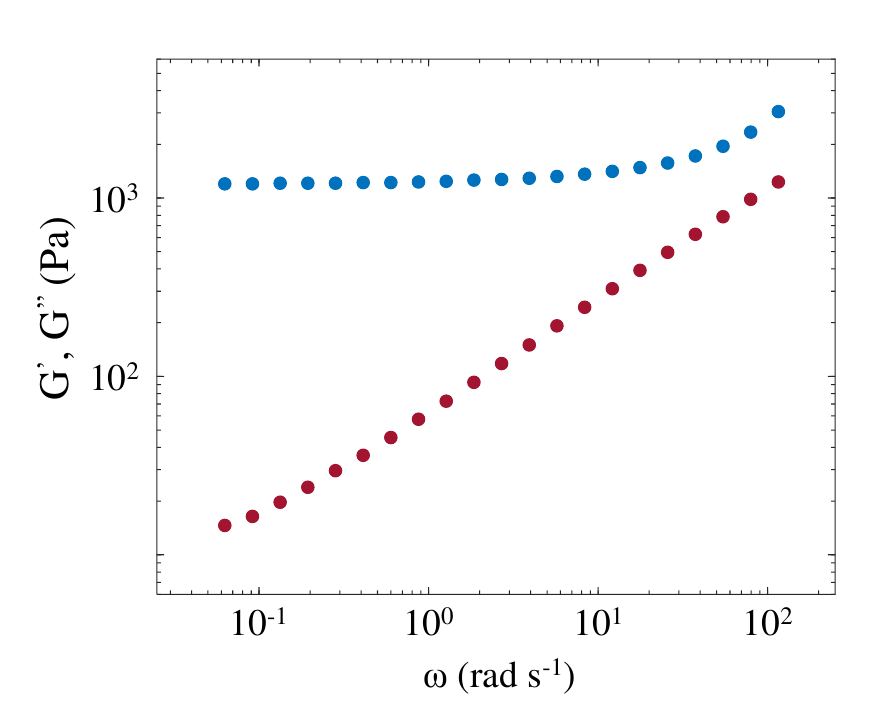}
\caption{\label{figS:Rheology} Rheology of the silicone gel. Blue dots correspond to the storage modulus and the red dots correspond to loss modulus. Silicone gel rheology was characterized by small amplitude oscillatory shear measurements by a strain-controlled rheometer (Physica MCR 500; Anton Paar, Austria) with a parallel plate geometry (PP20-MRD) at controlled temperature, 25$\pm$0.2$^{\circ}$C. The gap was set at 0.6 mm and the strain was fixed as 1$\%$.  The gel sample was directly crosslinked onto the rheometer plate.}
\end{figure}

\bibliography{apssamp}

\end{document}